\documentclass[aps,prl,twocolumn,showpacs]{revtex4-1}
\usepackage{amsmath}
\usepackage{amssymb}
\usepackage{graphicx}
\usepackage{epstopdf}

\DeclareMathOperator{\sech}{\mathrm{sech}}

\begin{document}
\title{Thermally-Activated Phase Slips in Superfluid Spin Transport in Magnetic Wires}

\author{Se Kwon Kim}
\affiliation{
	Department of Physics and Astronomy,
	University of California,
	Los Angeles, California 90095, USA
}

\author{So Takei}
\affiliation{
	Department of Physics and Astronomy,
	University of California,
	Los Angeles, California 90095, USA
}

\author{Yaroslav Tserkovnyak}
\affiliation{
	Department of Physics and Astronomy,
	University of California,
	Los Angeles, California 90095, USA
}

\date{\today}

\begin{abstract}
We theoretically study thermally-activated phase slips in superfluid spin transport in easy-plane magnetic wires within the stochastic Landau-Lifshitz-Gilbert phenomenology, which runs parallel to the Langer-Ambegaokar-McCumber-Halperin theory for thermal resistances in superconducting wires. To that end, we start by obtaining the exact solutions for free-energy minima and saddle points. We provide an analytical expression for the phase-slip rate in the zero spin-current limit, which involves detailed analysis of spin fluctuations at extrema of the free energy. An experimental setup for a magnetoeletric circuit is proposed, in which thermal phase slips can be inferred by measuring nonlocal magnetoresistance.
\end{abstract}


\maketitle

\emph{Introduction.}|A wire can carry an electrical current without dissipation under favorable conditions in the superconducting state, which is characterized by a complex-valued function of position, $\Psi(\mathbf{r})$, referred to as the superconducting order parameter describing a condensate of constituent particles \cite{Tinkham2004}. When density fluctuations of the condensate are energetically suppressed and thus the magnitude of the order parameter is constant, an electrical supercurrent is proportional to the gradient of the phase of the order parameter. In some circumstances, e.g., for thin wires or in the presence of strong magnetic fields, finite resistances arise \cite{HalperinIJMPB2010}, of which understanding has required both theoretical and experimental efforts to be made over the last decades. In particular, the theory for intrinsic thermal resistances in thin superconducting wires has been pioneered by \textcite{LittlePR1967} who suggested that dissipation occurs via thermally-activated phase slips (TAPS), jumps of the phase winding along the wire by $2 \pi$, that necessarily accompanies quenching of the order parameter at some point in the wire. The quantitative theory for TAPS has been developed by \textcite{LangerPR1967} and further elaborated by \textcite{McCumberPRB1970}, and has, therefore, been given the acronym following their initials: LAMH theory. 

No superconductivity has been observed at room temperature, challenging its practical utilization, whereas magnetism---another phenomenon resulting from spontaneous ordering---is ubiquitous in nature even at elevated temperatures. Being integrated with the information processing technology, it has spawned the field of spintronics \cite{WolfScience2001, *ZuticRMP2004}. A spin analog of an electrical supercurrent, superfluid spin transport, has been proposed in magnets with easy-plane anisotropy, where the direction of the magnetic order parameter within the easy plane plays a role of the phase of the superfluid order parameter \cite{SoninJETP1978, *SoninAP2010, KonigPRL2002, TakeiPRL2014, *TakeiPRB2014, ChenPRB2014}. Here, dissipationless spin current (polarized out of the easy plane) is sustained by a planar spiraling texture of the magnetic order. The absence of strict conservation laws for spin, e.g., due to Gilbert damping, rules out faithful analogy to electrical supercurrent, which requires us to consider superfluid spin transport as being distinct from conventional charge superfluid.

In this Letter, we theoretically study TAPS in superfluid spin transport in easy-plane magnetic wires within the Landau-Lifshitz-Gilbert (LLG) phenomenology. In equilibrium, the magnetic order is kept in the easy plane and, thus, can be characterized by its winding number, the total azimuthal-angle (phase) change along the wire. At a finite temperature, the winding number can increase or decrease due to thermal spin fluctuations via events that can be identified as TAPS. The most probable path for the dynamics of the magnetic order parameter during TAPS traverses the saddle point of the free energy, where a few localized spins develop significant out-of-easy-plane components. We obtain the exact solution for these saddle points by solving the time-independent Landau-Lifshitz equation. We also provide an analytical expression for the rate of TAPS in the zero spin-current limit, which involves detailed analysis of spin fluctuations at extrema of the free energy. To observe TAPS in magnetic wires, we adopt a magnetoeletric circuit proposed in Ref.~\cite{TakeiarXiv2015}, in which detection of nonlocal magnetoresistance can yield signatures of TAPS.

\begin{figure}
\includegraphics[width=\columnwidth]{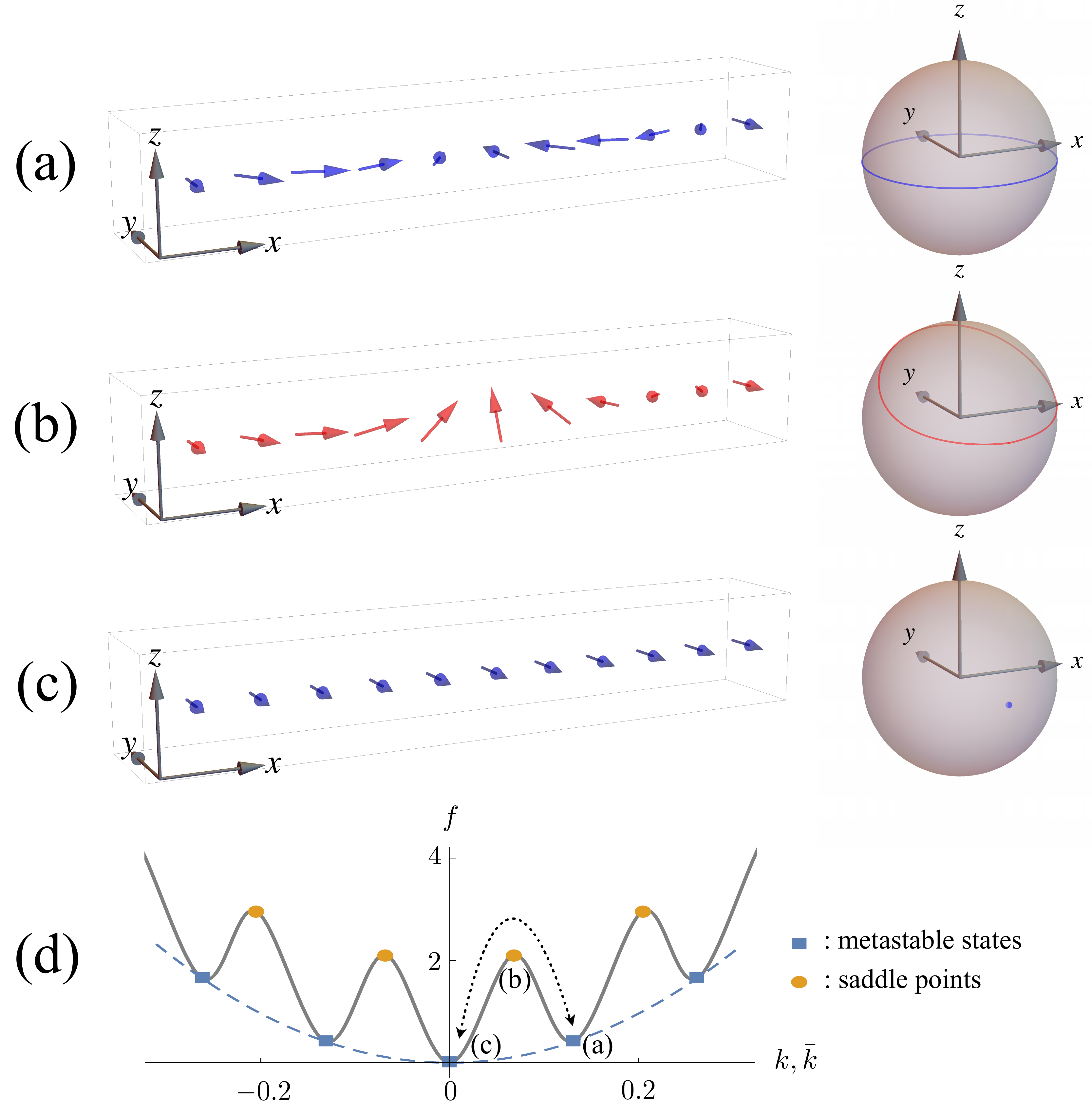}
\caption{(color online) (a)-(c) Extrema of the free energy $f$ [Eq.~(\ref{eq:f})] for an easy-$xy$-plane magnetic wire with periodic boundary conditions. The metastable state (a) that winds in the easy plane once will decay to the ground state (c) via TAPS passing over the saddle point (b), where a few spins localized within the magnetic coherence length develop significant out-of-easy-plane components. (d) A plot of the free energy $f$ as a function of spin current $k$ (metastable states) and $\bar{k}$ (saddle points) for the wire length $l = 48$. A solid line is a guide to the eye. A dashed line shows the free energy of the metastable states for an infinitely long wire: $f = k^2 l / 2$. Points corresponding to the three configurations, (a)-(c), are denoted accordingly. A dotted line illustrates transitions between nearby metastable states (a) and (c).}
\label{fig:fig1}
\end{figure}

\emph{Main results.}|We consider a thin easy-$xy$-plane magnetic wire with free energy $F[\mathbf{n}] \equiv \int dV ( A |\boldsymbol{\nabla} \mathbf{n}|^2 + K n_z^2 ) / 2$, where positive constants $A$ and $K$ parametrize the stiffness of the order parameter and the easy-plane anisotropy, respectively. Here, the unit vector $\mathbf{n} (x)$ is the direction of the order parameter: the local spin angular-momentum density for ferromagnets and the local N\'eel order for antiferromagnets. When the wire is narrow compared to the magnetic coherence length $\xi \equiv \sqrt{A/K}$, variations of the order parameter across the wire (of cross section $S$) can be neglected, which allows us to treat the order parameter $\mathbf{n}$, at a given time, as a function of position $x$ along the wire \cite{KosevichPR1990}. It is convenient to parametrize $\mathbf{n}$ in spherical coordinates, $\theta$ and $\phi$, defined by $\mathbf{n} \equiv (\sin \theta \cos \phi, \sin \theta \sin \phi, \cos \theta)$, with a rescaled free energy:
\begin{equation}
\label{eq:f}
f [\theta, \phi] \equiv \int_{-l/2}^{l/2} dx [ \theta'^2 + \sin^2 \theta \phi'^2 + \cos^2 \theta ] / 2 \,
\end{equation}
measured in units of $F_0 \equiv S \xi K$ (which is the maximum anisotropy energy that can be stored within the coherence length), where the position variable $x$ is measured in units of $\xi$ and runs over the wire length $l \equiv L/\xi$.

Extrema of the free energy are solutions of the time-independent Landau-Lifshitz equation:
\begin{subequations}
\label{eq:dU}
\begin{align}
\delta f / \delta \theta 	&= - \theta '' + \sin \theta \cos \theta \phi'^2 - \sin \theta \cos \theta = 0 \, , \label{eq:dU1} \\
\delta f / \delta \phi 	&= - (\sin^2 \theta \phi')' = 0 \, . \label{eq:dU2}
\end{align}
\end{subequations}
The second equation is the consequence of the invariance of the free energy under spin rotations about the $z$ axis. For static configurations, the associated conservation law describes spatial independence of the $z$-component of the spin current, $I_s \equiv - A S \sin^2 \theta \phi'$, so that the dimensionless constant parameter $k \equiv - I_s \xi / A S$ can be used to index solutions of Eqs.~(\ref{eq:dU}). There are two types of solutions of interest to us. The first is a local minimum of the free energy:
\begin{equation}
\theta (x) = \pi/2 \, , \quad \phi (x) = \phi_0 + k x \quad (|k| < 1) \, ,
\label{eq:m}
\end{equation}
with $\phi_0$ an arbitrary reference angle. There is a critical current, $|k| = 1$, for stable superfluid spin transport according to the Landau criterion \cite{LandauZETF1941, SoninJETP1978, KonigPRL2002}, above which spin fluctuations destabilize superfluidity. When the wire is long enough $l \gg 1$ (which is assumed henceforth), actual boundary conditions at the ends of the wire are not important. Imposing periodic boundary conditions on the order parameter, $\mathbf{n}(x = -l/2) = \mathbf{n}(x = l/2)$, quantizes the total azimuthal-angle change: $\Delta \phi \equiv \phi(l/2) - \phi(-l/2) =  2 \pi \mu$, in terms of integer $\mu$. The allowed values of $k$ are thus $k_\mu = 2 \pi \mu / l$. Figures~\ref{fig:fig1}(a) and~\ref{fig:fig1}(c) show the free-energy minima with winding numbers $\mu = 1$ and $\mu = 0$, respectively. 

At zero temperature, thermal spin fluctuations are frozen out. Persistent spin current in a closed magnetic ring, therefore, can be sustained indefinitely, when disregarding quantum spin fluctuations \cite{BraunPRB1997}. Finite temperature, however, agitates spins and opens transition channels between the metastable states carrying different spin current [see a dotted line in Fig.~\ref{fig:fig1}(d)]. The total azimuthal-angle change $\Delta \phi = 2 \pi \mu$ is quantized and well defined provided that the order parameter $\mathbf{n}$ avoids the poles, $|n_z| = 1$, where the azimuthal angle $\phi$ is ambiguous. In continuous transitions between two minima with different winding numbers, $\mu \neq \mu'$, the order parameter must hit one of the poles; this is analogous to the vanishing of the superconducting order parameter during TAPS \cite{LittlePR1967}. Supposing $T \ll F_0$, the transitions between metastable states are rare, which we assume throughout.

The most probable path of the order parameter during the transition between two metastable states will pass over the intervening saddle point of the free energy \cite{McCumberPRB1970, HalperinIJMPB2010}, which is the second kind of solution of Eq.~(\ref{eq:dU}) that we obtain with spatially varying $\bar{\theta} (x)$ \footnote{The solution in Eqs.~(\ref{eq:s}) corresponds to a single TAPS event located at $x = 0$, in which $\bar{\theta} (x)$ is a monotonic function of $x$ for $x < 0$. There are other saddle-point solutions describing simultaneous and multiple TAPS events. We only consider saddle points describing a single TAPS event for energetic considerations in this Letter.}:
\begin{subequations}
\label{eq:s}
\begin{align}
& \bar{\theta} (x)  	= \cos^{-1} \left[ \sqrt{1 - \bar{k}^2} \sech(\sqrt{1 - \bar{k}^2} x) \right] \, , \\
& \bar{\phi} (x) 			= \phi_0 + \bar{k} x + \tan^{-1} \left[ \frac{\sqrt{1 - \bar{k}^2} \tanh(\sqrt{1 - \bar{k}^2} x)}{\bar{k}} \right] \, ,
\end{align}
\end{subequations}
indexed by spin current $\bar{k}$, and any spatial translation thereof. This exact saddle-point solution constitutes our first main result. Periodic boundary conditions on $\mathbf{n}$ discretize allowed values of $\bar{k}$: $\Delta \phi = \bar{k}_\mu l + 2 \tan^{-1} [ (1 - \bar{k}_\mu^2)^{1/2} / \bar{k}_\mu ] = 2 \pi \mu$, where the quantities exponentially small for large $l$ are ignored here and hereafter. Figure~\ref{fig:fig1}(b) depicts the saddle-point solution with $\mu = 1$, which mediates the transition between two minima with $\mu = 1$ and $\mu = 0$. The spin currents of the metastable states and the saddle-point solutions interlace: $k_{\mu - 1} < \bar{k}_\mu < k_\mu$ (for positive $\mu$), meaning that there always exists the unique saddle point between two nearest metastable states. See Fig.~\ref{fig:fig1}(d) for an illustration.

The rate of transitions, respectively increasing or decreasing spin-current magnitude, may be written in the form
\begin{equation}
\label{eq:Gamma}
\Gamma_\pm = \Omega e^{- \Delta F_\pm / T} \, ,
\end{equation}
where temperature is measured in energy units so that $k_B = 1$. Here, $\Delta F_\pm \equiv F_0 \cdot \Delta f_\pm$ is the free-energy barrier to reach the intermediate saddle point, and $\Omega$ is the prefactor that depend on details of spin fluctuations around the extrema \footnote{The difference in the prefactors $\Omega$ for transitions that increase and decrease spin-current magnitude can be neglected in the regime $k \ll 1$ \cite{HalperinIJMPB2010}.}. Specifically, for the transitions between the two metastable states [Eq.~(\ref{eq:m})] with $k_{\mu}$ and $k_{\mu - 1}$ via the saddle point [Eq.~(\ref{eq:s})] with $\bar{k} = \bar{k}_\mu > 0$, the free-energy barriers can be directly obtained by evaluating the differences in the free energy $f$ [Eq.~(\ref{eq:f})]:
\begin{subequations}
\begin{align}
\Delta f_- (\bar{k}) 	&= 2 \sqrt{1 - \bar{k}^2} - 2 \bar{k} \tan^{-1} [ \sqrt{1 - \bar{k}^2} / \bar{k}] \, , \\
\Delta f_+ (\bar{k})	&= \Delta f_- (\bar{k}) + 2 \pi \bar{k} \, .
\end{align}
\end{subequations}
Since $\Delta f_- \le \Delta f_+$, fluctuations tend, on average, to reduce the spin-current magnitude and thus give rise to equilibriation. In the limit of zero current, $\bar{k} \rightarrow 0$, the free-energy barrier is $\Delta F \equiv 2 F_0 = 2 S \xi K$, which roughly represents the energy cost due to the out-of-easy-plane component of the order parameter in the phase slip region localized within the magnetic coherence length $\xi$.

Our second main result, which is derived in the supplemental material \footnote{See the supplemental material for details of the derivation.}, is the analytical expression of the prefactor $\Omega$ for ferromagnets in the zero spin-current limit:
\begin{equation}
\label{eq:main2}
\Omega (T) = \frac{1}{\pi \sqrt{2 \pi}} \frac{\alpha K}{ (1 + \alpha^2) s} \frac{L}{\xi} \sqrt{ \frac{\Delta F}{T} } \, ,
\end{equation}
which is analogous to the result for the superconducting wire in the LAMH theory \cite{McCumberPRB1970}, where $\alpha$ is the Gilbert damping constant and $s$ is the local spin angular momentum density. Here, $\alpha K / (1 + \alpha^2) s$ is the inverse of the relaxation time for the perturbed uniform easy-plane ferromagnet to return to the equilibrium state; $L/\xi$ represents the number of possible independent phase-slip locations; $\sqrt{\Delta F / T}$ stems from the breaking of the translational invariance of the system by the saddle point \footnote{When a saddle point breaks $m$ independent continuous symmetries that are respected by a metastable state, the associated transition rate is proportional to $(\Delta F / T)^{m/2}$ \cite{BraunAP2012}.}. The prefactor for antiferromagnets on bipartite lattice can be obtained by replacing $\alpha K / (1 + \alpha^2) s$ with $K / \alpha s$ for overdamped dynamics \footnote{The antiferromagnetic LLG equation is second order in time derivative \cite{AndreevSPU1980, *KimPRB2014} in general, whereas the theory of nucleation rates \cite{LangerPR1967} that we utilize requires that the equation is linear order in time derivative. For overdamped dynamics (when the Gilbert damping constant is larger than the ratio of the lattice constant to the coherence length, $\alpha \gg a / \xi$), the second-order term in the antiferromagnetic LLG equation is negligible, which allows us to employ the method in Ref.~\cite{McCumberPRB1970}.}, where $s$ is the local spin angular-momentum density per each sublattice.

\emph{Decay of persistent spin current.}|The persistent spin current in a closed ring will decay via TAPS at a finite temperature. From Eq.~(\ref{eq:Gamma}), the winding number $\mu = \Delta \phi / 2 \pi$, which characterizes metastable states, decays with the rate
\begin{subequations}
\label{eq:dr}
\begin{align}
\Gamma_+ - \Gamma_-  	&= - 4 \pi^2 (\xi F_0 / L T) \Omega(T) e^{- 2 F_0 / T} \mu \, , \\
					&\equiv - \kappa (T) \mu
\end{align}
\end{subequations}
to linear order in the winding number $\mu$ \footnote{In thermal equilibrium, the average energy of metastable states is of order of temperature: $F_0 k^2 l \sim T$. The condition justifying linearization, $\exp(\pi k F_0 / T) - \exp(- \pi k F_0 / T) \simeq 2 \pi k F_0 / T$, is $k F_0 / T \ll 1$, which is equivalent to $l \gg F_0 / T$ after $k$ is replaced by $\sqrt{T / F_0 l}$. Since the theory for nucleation rates \cite{LangerPR1967} that we depend on is based on the assumption that the energy barrier is much higher than the temperature, the parameter regime in which Eq.~(\ref{eq:dr}) is applicable is $l \gg F_0 / T \gg 1$.}. The spatially-averaged spin current $I_s \equiv 2 \pi \mu A S / L$ decays with the rate $\kappa(T) I_s$. Note that $\kappa(T)$ is independent of the length of the wire since $\Omega(T) \propto L$.

The dissipation of the spin current dictates the presence of the effective random force on the spin current to meet the fluctuation-dissipation theorem \cite{LL5}. The resultant stochastic dynamics of the spin current is described by
\begin{equation}
\dot{I}_s (t) = - \kappa(T) I_s (t) + \nu (t),
\end{equation}
where the white-noise Langevin term $\nu(t)$ with the correlator $\langle \nu(t) \nu(t') \rangle = 2 (A S / L) \kappa(T) T \delta (t - t')$ is introduced to yield the thermal variance of the spin current, $\langle I_s^2 \rangle = (A S / L) T$, which we obtain from the thermal expectation value of the free energy.

\begin{figure}
\includegraphics[width=\columnwidth]{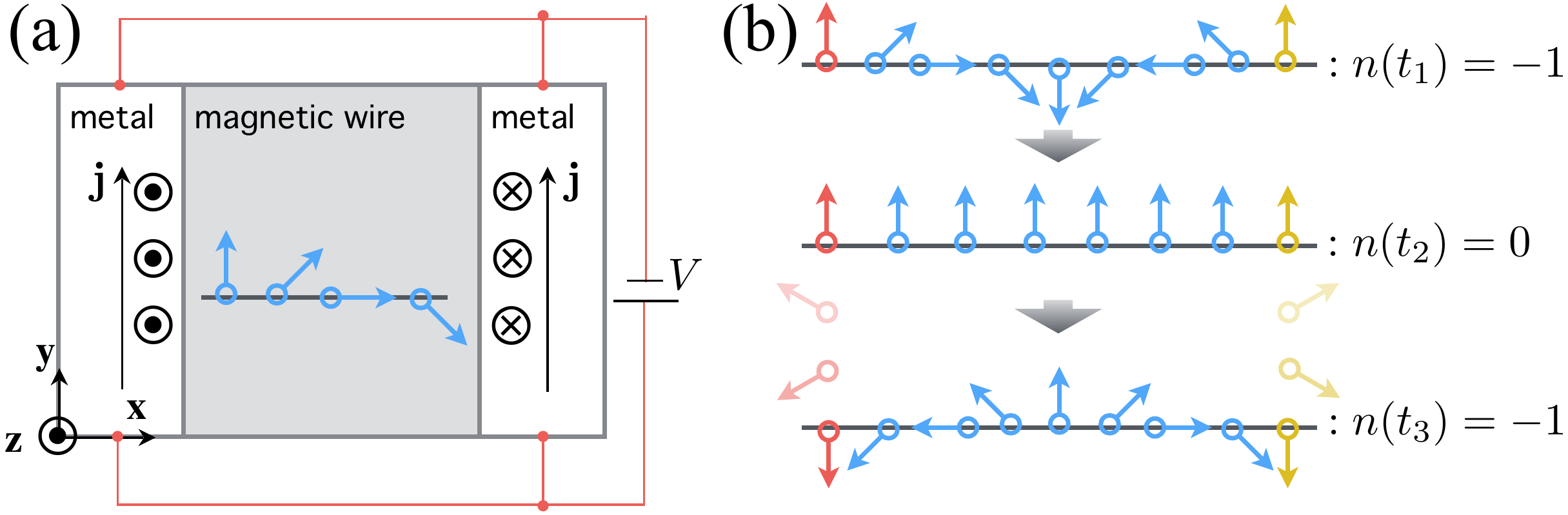}
\caption{(color online) (a) Schematics of an experimental setup for detecting TAPS, in which two identical metals, parallel in the electric circuit, are connected by a magnetic insulator supporting superfluid spin transport. (b) Schematics illustrating the origin of an electromotive force in the metals. TAPS unwind the equilibrium spiraling structure (at $t = t_1$), resulting in the uniform state (at $t = t_2$). As the magnet returns to the equilibrium spiraling structure, the magnetization at the left (right) interface rotates counterclockwise (clockwise), which in turn induces a detectable electromotive force in the metals.} 
\label{fig:fig2}
\end{figure}

\emph{Discussion.}|TAPS in superfluid spin transport can be detected in an experimental setup proposed in Ref.~\cite{TakeiarXiv2015}, in which two identical metals connected parallel in the external electric circuit are linked by a thin easy-plane magnetic insulating wire (see Fig.~\ref{fig:fig2}). In the presence of spin-orbit coupling at metal\textbar magnet interface, current in the metal gives rise to a torque in the magnet, and, as an Onsager reciprocal effect, dynamics of magnetic moments induces an electromotive force in the metal \cite{TserkovnyakPRB2014}. 

At zero temperature, this configuration supports static spiraling structure of the magnetization \cite{TakeiarXiv2015}, with the left metal injecting and the right metal draining spin current [see Fig.~\ref{fig:fig2}(a)]. The associated electromotive force is absent, and the effective resistivity of the circuit, therefore, is not affected by spin superfluid. At a finite temperature, however, TAPS unwind the spiraling structure stochastically with the net rate of $\kappa(T) \mu(t)$, where $\mu(t)$ is the winding number at fixed time $t$. As the magnetization rewinds to the equilibrium spiraling structure, the magnetic moment at the left (right) interface rotates counterclockwise (clockwise), which induces an electromotive force in the adjacent metals \footnote{Being localized in the bulk of the wire, TAPS do not affect the magnetic moments at the interfaces, for a wire that is sufficiently long.}. In the steady state, the time-averaged precession of the magnetization at the interfaces are $\dot{\phi}_l (t) = - \dot{\phi}_r (t) = - \pi \kappa(T) \mu(t)$. Induced electromotive force reduces the effective resistivity of the circuit (following the derivation of Ref.~\cite{TakeiarXiv2015}): $\rho \rightarrow \rho + \rho_m$ with $\rho_m = - \vartheta^2 / [\hbar g^{\uparrow \downarrow} / 4 \pi + 2 A / \kappa(T) L]$, where $\rho$ is the resistivity of the metal, $g^{\uparrow \downarrow}$ is the effective interfacial spin-mixing conductance (parametrizing the Gilbert-damping enhancement at the interfaces due to spin pumping), and $\vartheta$ is related to the effective interfacial spin Hall angle $\Theta$ via $\vartheta \equiv (\hbar / 2 e t) \tan \Theta$, with $t$ being the thickness of the metals in the $x$ direction \cite{TakeiarXiv2015}. Observation of the characteristic dependence of the effective resistivity of the circuit on the length of the wire (algebraic) or temperature (exponential) would provide experimental signatures of TAPS.

Spins are treated classically in our theory for TAPS. Quantum aspect of spins would become important in the low-temperature regime, where quantum phase slips may become a dominant source of dissipation of spin supercurrent. It is possible to pursue a study of quantum phase slips with the aid of semiclassical quantization of spins, which has been used to understand macroscopic quantum tunneling of magnetic domain walls \cite{BraunPRB1997}. Recalling about the superconductor-to-insulator quantum phase transition (as a function of wire cross section) \cite{SchmidPRL1983, HalperinIJMPB2010}, a natural question arises on the role of the spin Berry phase in the analogous physics of the magnetic quantum phase slips.

\begin{acknowledgments}
This work was supported by the Army Research Office under Contract No. 911NF-14-1-0016 and in part by the U.S. Department of Energy, Office of Basic Energy Sciences under Award No. DE-SC0012190 and FAME (an SRC STARnet center sponsored by MARCO and DARPA).
\end{acknowledgments}

\bibliographystyle{/Users/evol/Dropbox/School/Research/apsrev4-1-nourl}
\bibliography{/Users/evol/Dropbox/School/Research/master}

\end{document}


\title{Supplemental Material: Thermally-Activated Phase Slips in Superfluid Spin Transport in Magnetic Wires}

\author{Se Kwon Kim}
\affiliation{
	Department of Physics and Astronomy,
	University of California,
	Los Angeles, California 90095, USA
}

\author{So Takei}
\affiliation{
	Department of Physics and Astronomy,
	University of California,
	Los Angeles, California 90095, USA
}

\author{Yaroslav Tserkovnyak}
\affiliation{
	Department of Physics and Astronomy,
	University of California,
	Los Angeles, California 90095, USA
}

\date{\today}

\maketitle

In this supplemental material, we derive the prefactor $\Omega(T)$ in the phase-slip rates in the limit of zero spin current.

\emph{Phase-slip rates.}|To derive the prefactor $\Omega(T)$, it is convenient to impose antiperiodic boundary conditions on the order parameter: $\mathbf{n} (x = -l/2) = - \mathbf{n} (x = l/2)$, with which a unique saddle point solution is present at $\bar{k} = 0$. There are two degenerate ground states with the total azimuthal angle changes $\Delta \phi = \pm \pi$. The saddle point passed during the transition between these two states is analogous to Eq.~(4) with $\bar{k} = 0$:
\begin{equation}
\label{eq:s0}
\bar{\theta} = \cos^{-1} (\sech x) \, , \quad \bar{\phi} = \phi_0 + (\pi/2) H(x) \, ,
\end{equation}
where $H(x)$ is the Heaviside step function.

For the probability per unit time of a single TAPS event, adapting the derivation for superconducting wires in Ref.~\cite{McCumberPRB1970} to our problem, we obtain the following expression for general spin current $k$:
\begin{equation}
\Gamma(k, T) = \Omega (k, T) e^{- \Delta F (k) / T} \, ,
\end{equation}
with the prefactor
\begin{equation}
\label{eq:Omega}
\Omega (k, T) = 	\frac{\gamma (k) \Lambda (k) }{2 \pi} \sqrt{ \frac{F_0}{2 \pi T} }
				\sqrt{ \frac{\det' \mathcal{H}}{|\det' \bar{\mathcal{H}}|} } \, .
\end{equation}
Similar expression has been derived in Refs.~\cite{BraunPRL1993, *BraunPRB1994} to obtain rates of magnetization reversal in ferromagnetic wires with easy-axis anisotropy, where transitions between two ground states with uniform magnetization are driven by soliton-antisoliton nucleations. Here, $\gamma(k)$ is the escape rate from the involved saddle point, according to the linearized equation of motion at the saddle point solved by $\delta \mathbf{n} (x, t) = \exp(\tilde{\gamma} t) \delta \mathbf{n} (x)$ with $\gamma = \Re{\tilde{\gamma}}$ \footnote{$\gamma$ is the only quantity that requires the dynamical equation of $\mathbf{n}$. For its calculation, we follow Ref.~\cite{BraunPRL1993}. All other quantities, e.g., $\Lambda$, can be derived from the (static) free energy.}; $\Lambda (k)$ stems from the presence of the zero mode associated with the translation symmetry breaking by the saddle point; $\mathcal{H}$ and $\bar{\mathcal{H}}$ (which are discussed further below) are the differential operators accounting for contributions from fluctuations $\delta \mathbf{n}$ to the free energy $f$ at the metastable states and saddle-point solutions, respectively; $\det' \mathcal{O}$ is the product of nonzero eigenvalues of the operator $\mathcal{O}$. 

\begin{figure}
\includegraphics[width=\columnwidth]{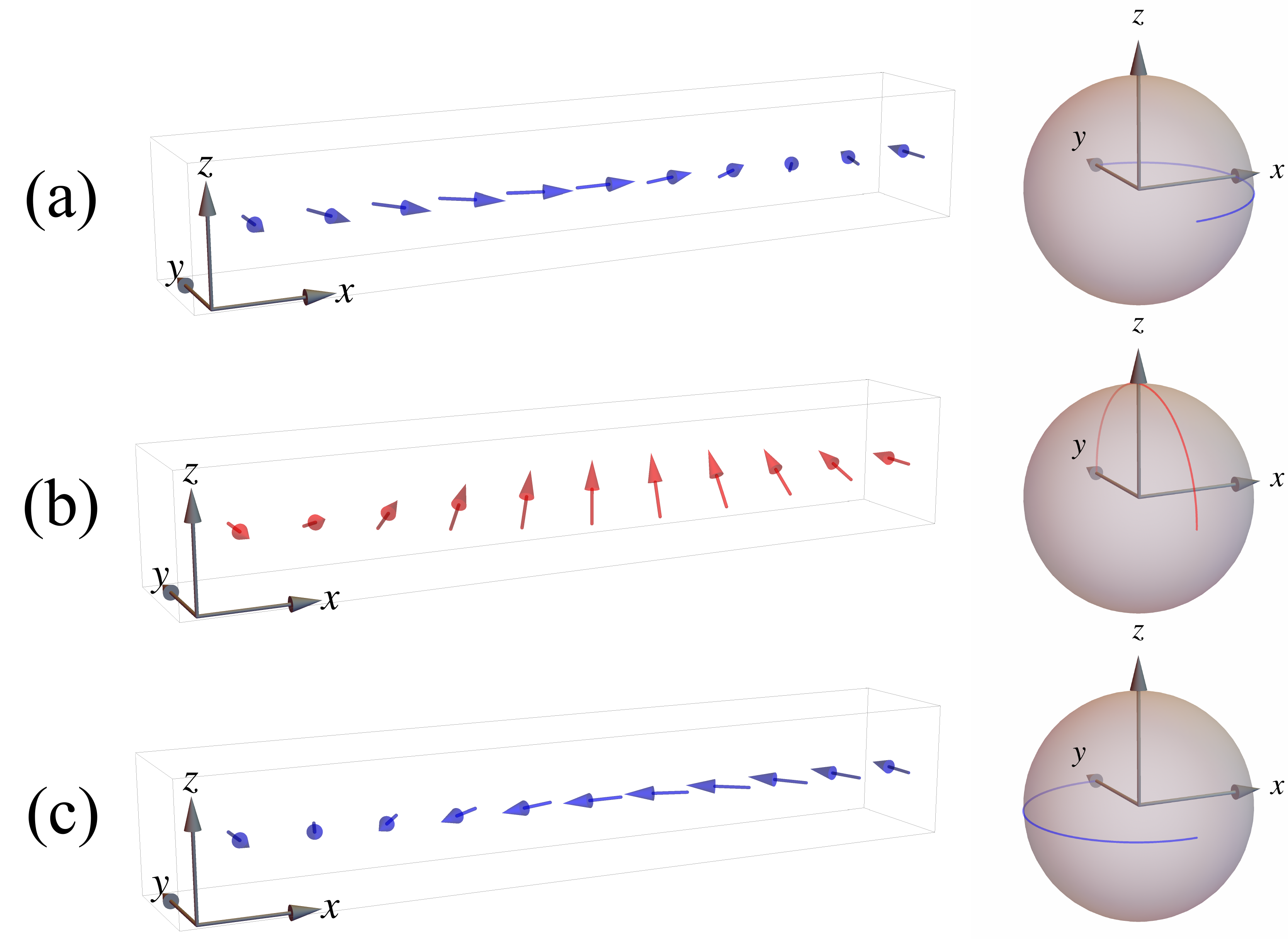}
\caption{(color online) Extrema of the free energy [Eq. (1)] with antiperiodic boundary conditions. Transitions between the two metastable states, (a) and (c), occur via TAPS passing over the saddle point (b), where a few spins in the phase slip region localized within the magnetic coherence length develop significant out-of-easy-plane components.}
\label{fig:figs1}
\end{figure}

\emph{Fluctuations.}|Let us investigate spin fluctuations at extrema of the free energy. To describe small deviations $\delta \mathbf{n}$ from the extrema $\mathbf{n}_e$, it is convenient to setup the local orthonormal frame defined by $\hat{\mathbf{e}}_3 \equiv \mathbf{n}_e, \, \hat{\mathbf{e}}_1 \equiv \partial \mathbf{n}_e / \partial \theta, \, \hat{\mathbf{e}}_2 \equiv (\sin \theta)^{-1} \partial \mathbf{n}_e / \partial \phi$, in which the deviation is parametrized by two numbers $u$ and $v$: $\mathbf{n} = u \hat{\mathbf{e}}_1 + v \hat{\mathbf{e}}_2 + (1 - u^2 - v^2)^{1/2} \hat{\mathbf{e}}_3$. 

The contribution from spin fluctuations, $\Psi(x) \equiv [u(x), v(x)]$, to the free energy $f$ at the metastable state with the spin current $k$ is $\int dx \, \Psi(x)^\mathrm{T} \mathcal{H} \Psi(x) / 2$, where
\begin{equation}
\mathcal{H} \equiv \begin{pmatrix} - d^2/dx^2 + 1 - k^2 & 0 \\ 0 & - d^2/dx^2 \end{pmatrix} \, .
\end{equation}
The corresponding eigenvalue problems for $u(x)$ and $v(x)$ are decoupled:
\begin{subequations}
\label{eq:fm}
\begin{align}
(- d^2/dx^2 + 1 - k^2) u_n (x) &= \epsilon_n u_n(x) \, , \\
(- d^2/dx^2) v_n (x) &= \epsilon'_n v_n (x) \, .
\end{align}
\end{subequations}
Note that above the critical current, $|k| = 1$, the stable spin current cannot be sustained due to unrestricted spin fluctuations. 

Similarly, the contribution from spin fluctuations to the free energy at the saddle point with $k = 0$ is $\int dx \, \bar{\Psi}(x)^\mathrm{T} \bar{\mathcal{H}} \bar{\Psi}(x) / 2$, where
\begin{equation}
\bar{\mathcal{H}} \equiv \begin{pmatrix} - d^2/dx^2 + 1 - 2 \sech^2 x & 0 \\ 0 & - d^2/dx^2 - 2 \sech^2 x \end{pmatrix} \, .
\end{equation}
The eigenvalue problem for $\bar{u}(x)$ and $\bar{v}(x)$ also decouples \footnote{For general spin current $\bar{k} \neq 0$, $\bar{u} (x)$ and $\bar{v} (x)$ are coupled in the eigenvalue problem, which does not allow simple solutions.}:
\begin{subequations}
\label{eq:fs}
\begin{align}
(- d^2/dx^2 + 1 - 2 \sech^2 x) \bar{u}_n (x) &= \bar{\epsilon}_n \bar{u}_n(x) \, , \\
(- d^2/dx^2 - 2 \sech^2 x) \bar{v}_n (x) &= \bar{\epsilon}'_n \bar{v}_n (x) \, .
\end{align}
\end{subequations}

Two eigenvalue problems, Eqs.~(\ref{eq:fm}) and~(\ref{eq:fs}), at $k = 0$ are related by the supersymmetry \cite{CooperPR1995, KimPRB2014} and are exactly solvable \footnote{The operators for $u(x)$ and $\bar{u}(x)$ at zero spin current, $\mathcal{H}_u \equiv - d^2 / dx^2 + 1$ and $\bar{\mathcal{H}}_u \equiv -d^2 / dx^2 +1 - 2 \sech^2 x$, can be factorized in terms of the same operators $a$ and $a^\dagger$: $\mathcal{H}_u = a a^\dagger$ and $\bar{\mathcal{H}}_u = a^\dagger a$, where $a \equiv d / dx + \tanh x$ and $a^\dagger \equiv - d/dx + \tanh x$, and are, thus, called supersymmetric partners. The operator $\mathcal{H}_u$ has plane-wave eigenstates: $u_q (x) = \exp(i q x)$. Continuum eigenstates of $\bar{\mathcal{H}}_u$ can be obtained from those of $\mathcal{H}_u$ by applying the operator $a^\dagger$: $\bar{u}_q (x) = a^\dagger u_q (x) = (\tanh x - i q) \exp(i q x)$. Furthermore, they have the same eigenvalues: $\mathcal{H}_u u_q (x) = \bar{\mathcal{H}}_u \bar{u}_q (x) = 1 + q^2$. There is a similar relation for $v(x)$ and $\bar{v}(x)$.}. Two properly-normalized localized solutions of Eq.~(\ref{eq:fs}) are of particular interest to us:
\begin{subequations}
\label{eq:ls}
\begin{align}
& \bar{\epsilon}'_1 = - 1, 	& \bar{v}_1 (x) = 2^{-1/2} \sech x \, , \\
& \bar{\epsilon}_1 = 0,	& \bar{u}_1 (x) = 2^{-1/2} \sech x \, .
\end{align}
\end{subequations}
The negative-eigenvalue solution $\bar{v}_1 (x)$ represents the direction of deviation $\delta \mathbf{n}$ from the saddle point that decreases the free energy the most rapidly, which shall be used to derive $\gamma$ in Eq.~(\ref{eq:Omega}). The zero-eigenvalue solution $\bar{u}_1 (x)$ is the Goldstone mode associated with the freedom of choice for a phase-slip location, which shall be used to derive $\Lambda$ in Eq.~(\ref{eq:Omega}).

\emph{Derivation of the prefactor.}|Let us now derive the escape rate from the saddle point, $\gamma$, in $\Omega(T)$ [Eq.~(\ref{eq:Omega})]. We will show details for the case of ferromagnets. The dynamics of $\mathbf{n}$ is described by the LLG equation, $s (1 +  \alpha \mathbf{n} \times) \dot{\mathbf{n}} = \mathbf{n} \times \mathbf{h}$, where $\mathbf{h} \equiv - \partial F / \partial \mathbf{n}$ is the effective field conjugate to $\mathbf{n}$. Writing $[\bar{u} (x, t), \bar{v} (x, t)] = \exp(\gamma t) [\bar{u} (x), \bar{v} (x)]$, the linearized LLG equation at the saddle point for $\bar{u} (x)$ and $\bar{v} (x)$ is given by
\begin{subequations}
\begin{align}
s \gamma [\bar{u} (x) - \alpha \bar{v} (x)] 	&= K (\bar{\mathcal{H}}_u - 1) \bar{v} (x) \, , \\
s \gamma [\bar{v} (x) + \alpha \bar{u} (x)] 	&= - K \bar{\mathcal{H}}_u \bar{u} (x) \, ,
\end{align}
\end{subequations}
with the differential operator $\bar{\mathcal{H}}_u \equiv - d^2/dx^2 + 1 - 2 \sech^2 x$. It is easy to verify that the equations are satisfied by the escape rate $\gamma = \alpha K / (1 + \alpha^2) s$ in conjunction with the eigenfunction $[\bar{u} (x), \bar{v} (x)] = \sech(x) \, [1, -\alpha]$ whose special profile was taken from Eq.~(\ref{eq:ls}). Applying the same method to antiferromagnets on bipartite lattice, whose dynamics is described by the antiferromagnetic LLG equation \cite{AndreevSPU1980, *KimPRB2014}, we obtain $\gamma = K / \alpha s$ for overdamped dynamics.

$\Lambda$ in Eq.~(\ref{eq:Omega}) can be derived following Ref.~\cite{McCumberPRB1970}. Infinitesimal spatial translations of the saddle-point solution [Eq.~(\ref{eq:s0})] can be represented by $d \bar{\theta} (x) = dx \, \sech(x)$. Equating $d \bar{\theta}(x)$ to $d \lambda \, \bar{u}_1 (x)$ ($\lambda$ parametrizing the associated zero-energy mode) yields $\Lambda = \int d\lambda = \int dx \sqrt{2} = \sqrt{2} l$.

For the ratio of the determinant $r \equiv \det' {\mathcal{H}} / |\det' \bar{\mathcal{H}}|$ in Eq.~(\ref{eq:Omega}), it is straightforward to generalize the method used by \textcite{McCumberPRB1970} for superconducting wires owing to the similarity between our eigenvalue problems [Eqs.~(\ref{eq:fm}) and~(\ref{eq:fs})] and their problems [Eqs.~(3.12) and (3.13) in the reference]. Applying the method of Ref.~\cite{McCumberPRB1970} with proper boundary conditions \footnote{One complication arises because of different boundary conditions. The basis vectors for spin deviations, $\hat{\mathbf{e}}_1 (x)$ and $\hat{\mathbf{e}}_2 (x)$, can change signs between $x = -l/2$ and $x = l/2$, depending on the extremum solution. To be consistent with the antiperiodic boundary condition on $\mathbf{n}$, periodic boundary conditions need to be imposed on $v(x)$ and $\bar{u}(x)$, and antiperiodic ones on $u(x)$ and $\bar{v}(x)$.} yields the exact value of the ratio: $r = 4$. This value can also be numerically verified by evaluating the expression with the exact eigenvalues that we are able to obtain with the aid of supersymmetry of Eqs.~(\ref{eq:fm}) and~(\ref{eq:fs}) \footnote{The eigenvalues of Eqs.~(\ref{eq:fm}) and~(\ref{eq:fs}) that appear in the ratio $r$ are $\epsilon(\mu) = 1 + [(2 \pi \mu + \pi) / l]^2, \epsilon'(\mu) = (2 \pi \mu / l)^2, \bar{\epsilon}(\mu) = 1 + \bar{q}_\mu^2, \bar{\epsilon}' (\mu) = (\bar{q}'_\mu)^2$, where $\mu$ is an integer. Here, $\bar{q}_\mu$ and $\bar{q}'_\mu$ are solutions of the equations: $\bar{q}_\mu l - 2 \tan^{-1}(\bar{q}_\mu) = 2 \pi \mu + \pi$ and $\bar{q}'_\mu l - 2 \tan^{-1}(\bar{q}'_\mu) = 2 \pi \mu$. For $l \gg 1$, the ratio 
\begin{equation}
\frac{
	\prod_{\mu=-m}^{m} \epsilon(\mu) \prod_{\mu = 1}^m \epsilon'(\mu) \prod_{\mu = -m}^{-1} \epsilon'(\mu)
}{
	 \prod_{\mu = -m+1}^m \bar{\epsilon}(\mu) \prod_{\mu = 1}^m \bar{\epsilon}'(\mu) \prod_{\mu=-m+1}^{-1} \bar{\epsilon}'(\mu)
}
\end{equation}
is verified to converge to $4$ as $m \rightarrow \infty$.}.

\bibliographystyle{/Users/evol/Dropbox/School/Research/apsrev4-1-nourl}
\bibliography{/Users/evol/Dropbox/School/Research/master}